\documentclass{elsart}
\usepackage{epsf}
\usepackage{amsmath}
\usepackage{cite}

\usepackage{amssymb}

\newcount\hour\newcount\minute
        \hour=\time \divide\hour by60 \minute=\time
        {\multiply\hour by60 \global\advance\minute by-\hour}
        \edef\militarytime{\number\hour:\ifnum\minute<10
0\fi\number\minute}

\makeatletter
\DeclareRobustCommand*\cal{\@fontswitch\relax\mathcal}
\makeatother

\makeatletter

\@addtoreset{equation}{section}

\def\draftdate{\relax}
\def\mda{\relax}
\def\mua{\relax}
\def\mla{\relax}
\def\draft{
\def\thtystars{******************************}
\def\sixtystars{\thtystars\thtystars}
\typeout{}
\typeout{\sixtystars**}
\typeout{* Draft mode!
         For final version remove \protect\draft\space in source file *}
\typeout{\sixtystars**}
\typeout{}
\def\draftdate{\today}
\def\mua{\marginpar[\boldmath\hfil$\uparrow$]%
                   {\boldmath$\uparrow$\hfil}%
                    \typeout{marginpar: $\uparrow$}\ignorespaces}
\def\mda{\marginpar[\boldmath\hfil$\downarrow$]%
                   {\boldmath$\downarrow$\hfil}%
                    \typeout{marginpar: $\downarrow$}\ignorespaces}
\def\mla{\marginpar[\boldmath\hfil$\rightarrow$]%
                   {\boldmath$\leftarrow $\hfil}%
                    \typeout{marginpar: $\leftrightarrow$}\ignorespaces}
\overfullrule 5pt
\oddsidemargin -15mm
\marginparwidth 29mm
}

\def\stars{\strut\leaders\hbox{*}\hfill\strut}
\def\starline{\hfil\strut\hfil\hbox to \textwidth {\stars}\hfil}

\newcommand\Ref[1]     {Ref.\,\cite{#1}}
\newcommand\Refs[1]    {Refs.\,\cite{#1}}
\newcommand\eqn[1]     {Eq.\,(\ref{#1})}
\newcommand\eqns[2]    {Eqs.\,(\ref{#1}) and~(\ref{#2})}

\newcommand\sect[1]    {Sect.\,{\ref{#1}}}

\newcommand\nn         {\nonumber}

\def\beq{\begin{equation}}
\def\eeq{\end{equation}}
\def\beeq{\begin{eqnarray}}
\def\eeeq{\end{eqnarray}}
\newcommand\bom[1]     {{\mbox{\boldmath $#1$}}}
\def\aand{&&}

\newcommand\as         {\ensuremath{\alpha_{\mathrm{s}}}}

\newcommand\gs         {\ensuremath{g_{\mathrm{s}}}}

\newcommand{\CA}       {C_{\mathrm{A}}}
\newcommand{\TR}       {T_{\mathrm{R}}}

\newcommand{\bT}       {\bom{T}}

\newcommand\qb         {{\bar q}}

\newcommand{\ep}       {\epsilon}        
\newcommand{\eps}      {\varepsilon}

\newcommand\ldot       {\!\cdot\!}

\newcommand{\ri}       {{\mathrm{i}}}

\newcommand\la         {\langle}
\newcommand\ra         {\rangle}

\newcommand{\cM}       {{\cal M}}
\newcommand\SME[3]     {|{\cal M}_{#1}^{(#2)}{(#3)}|^2}
\newcommand\M[2]       {\ensuremath{|{\cal{M}}_{#1}^{#2}|^2}}
\newcommand\bra[3]     {\la {\cal M}_{#1}^{#2}#3|}
\newcommand\ket[3]     {|{\cal M}_{#1}^{#2}#3\ra}

\newcommand{\bJ}[1]    {\bom{J}_{#1}}

\newcommand{\bC}[1]    {\bom{\mathrm C}_{#1}}

\newcommand{\bS}[1]    {\bom{\mathrm S}_{#1}}

\newcommand{\bSCS}[1]  {\bom{\mathrm C}\kern-2pt\bom{\mathrm S}_{#1}}

\def\hP{\hat{P}}
\newcommand{\calS}     {{\cal S}}

\newcommand{\cS}[2]    {{\cal S}_{#1}^{#2}}

\newcommand{\cSCS}[1]  {{\cal C}\kern-2pt{\cal S}_{#1}^{~}}

\newcommand{\pslash}   {p\kern-5pt/}

\begin{document}


\begin{frontmatter}
\begin{flushright}
hep-ph/0702273 \\
CERN-PH-TH/2006-231\\
ZU-TH 3/07
\end{flushright}

\title{Separation of soft and collinear infrared limits of QCD
squared matrix elements}
\author[CERN]{Zolt\'an Nagy}, 
\author[DE]{G\'abor Somogyi}
\author[UniZ]{and Zolt\'an Tr\'ocs\'anyi}\footnote{On leave from University
of Debrecen and Institute of Nuclear Research of the Hungarian Academy
of Sciences, Hungary.}
\address[CERN]{Physics Department, Theoretical Group, CERN, CH-1211 Geneva 23,
Switzerland}
\address[DE]{Institute of Nuclear Research of the Hungarian Academy of
Sciences, H-4001 Debrecen P.O.Box 51, Hungary}
\address[UniZ]{Institute for Theoretical Physics, University of
Z\"urich, Winterthurerstrasse 190, CH-8057 Z\"urich, Switzerland}
\date{\today}

\begin{abstract}
We present a simple way of separating the overlap between the soft and
collinear factorization formulae of QCD squared matrix elements. We
check its validity explicitly for single and double unresolved
emissions of tree-level processes. The new method makes possible the
definition of helicity-dependent subtraction terms for regularizing the
real contributions in computing radiative corrections to QCD jet cross
sections. This implies application of Monte Carlo helicity summation in
computing higher order corrections.
\end{abstract}

\end{frontmatter}


%
%

\section{Introduction}
\label{sec:intro}

The majority of available techniques for the computation of radiative
corrections in Quantum Chromodynamics (QCD) relies on the universal
factorization formulae of QCD matrix elements for the emission of
unresolved (soft and/or collinear) partons, which describe the
universal structure of the infrared singularities \cite{Altarelli:1977zs,%
Bassetto:1984ik}.  These factorization formulae have played an
essential role in devising completely general algorithms
\cite{Giele:1991vf,Giele:1993dj,Frixione:1995ms,Nagy:1996bz,%
Frixione:1997np,Catani:1996vz} for the cancellation of infrared
singularities when combining the tree-level and one-loop contributions
in the evaluation of jet cross sections at the next-to-leading order
(NLO) accuracy. More recently a new algorithm has been proposed
\cite{Somogyi:2006cz} that can also be extended to the computation of
the next order (NNLO) radiative corrections \cite{Somogyi:2006da,%
Somogyi:2006db}.  The essence of these algorithms is the definition of
suitable approximate matrix elements that have the same infrared
singular behaviour as the matrix elements themselves.  The
factorization formulae however, cannot be directly used as such
approximate matrix elements for two reasons.  On the one hand the
factorization formulae are unambiguously defined in the strict
unresolved limits. On the other hand the soft and collinear
factorization formulae overlap in regions of the phase space where the
unresolved parton is simultaneously soft and collinear to a hard
parton.  The naive application of the factorization formulae as
subtraction terms leads to double subtractions and as a result to
uncancelled infrared singularities.

There are two ways to cope with the double subtractions. One can either
introduce compensating terms for the double subtractions as for example
in the classic paper \cite{Ellis:1980wv}, or define single subtractions
that smoothly interpolate between the soft and collinear regions, as in
the dipole scheme \cite{Catani:1996vz}. Both techniques have also been
extended to NNLO computations. In \Refs{Campbell:1998nn,%
Gehrmann-DeRidder:2005cm,Daleo:2006xa} antennae subtraction terms have 
been introduced that smoothly interpolate among all doubly unresolved
regions, while the systematic way of accounting for the overlap among
the various factorization formulae for double unresolved emissions has
been worked out in \Ref{Somogyi:2005xz}. Thus the problem of double
subtractions could in principle be considered solved. Yet in this
letter we present a third solution to this problem, which is very
simple and can be extended to any order in perturbation theory easily.
It also offers some algorithmic advantages if one aims at automatizing
the computation of QCD radiative corrections. Furthermore, it allows
the introduction of helicity dependent subtraction terms as we show in
this paper.  

\section{Separation of the soft and collinear singularities at NLO}
\label{sec:softsep_NLO}

The separation of the soft and collinear singularities can be obtained
from the different physical picture of the two cases. In a physical
gauge the collinear singularities are due to the collinear splitting of
an external parton \cite{Curci:1980uw}. The overall colour structure of the
event does not change, the splitting is entirely described by the
Altarelli--Parisi functions which are a product of colour factors%
\footnote{Eigenvalues of the quadratic Casimir operators of the emitted
collinear gluon in the representation of the parent parton, or the
normalization factor of the colour charges in the case of gluon splitting
into fermions.}
and a kinematical function describing the collinear kinematics of the
splitting. The emission of a soft gluon is just the opposite. It does not
affect the kinematics (momenta and spins) of the radiating partons, but
it affects their colour because it always carries away some colour
charge. As a result it leads to colour correlations that can be
percieved as a soft gluon cloud around the event.
\begin{figure}
\centerline{
\epsfxsize=8cm \epsfbox{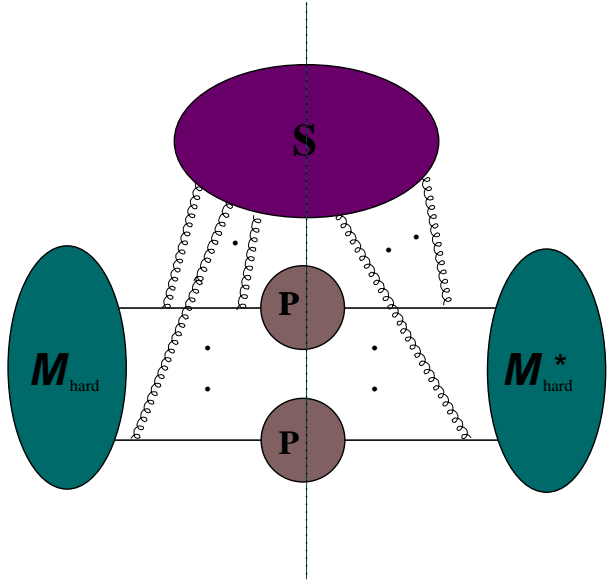}
}
\caption{General structure of infrared factorization at any perturbative
order. (From \Ref{Catani:1999ss} with permission of S. Catani.)}
\label{fig:general}
\end{figure}

The analytic expression for the soft factorization of the QCD matrix
elements can be derived using the soft-gluon insertion technique 
\cite{Bassetto:1984ik,Catani:1999ss}.
To describe this technique, we use the colour-state notation \cite{Catani:1996vz} for
the amplitudes and the operator notation \cite{Somogyi:2005xz} for taking the soft
limit of the amplitude. In the colour-state notation the amplitude for
$n$ external partons is represented by the ket vector
$\ket{n}{}{(p_1,\dots,p_n)}$. In the limit, when momentum $p_r^\mu$
becomes soft, the $m+1$ parton amplitude fulfills 
the following factorization formula
(for the precise meaning of the operator $\bS{r}$ and also that
of the operator $\bC{ir}$ used below, we refer to \Ref{Somogyi:2005xz})
\beq
\bS{r} \la c_r \ket{m+1}{}{(p_r,\dots)} = \eps^\mu(p_r)
\bJ{\mu}(r,\ep) \ket{m}{}{(\dots)}
\,,
\label{eq:SrM}
\eeq
where $c_r$ is the colour-index of parton $r$. Both the amplitude and the
soft current $J_r$ can be expanded in perturbation theory,
\beeq
\ket{}{}{} \aand= \ket{}{(0)}{} + \ket{}{(1)}{} + \dots\,,
\label{eq:Mexpansion}
\\[3mm]
\bJ{\alpha}(r, \ep) \aand= \gs \mu^\ep
\Big[\bJ{\alpha}^{(0)}(r) + (\gs \mu^\ep)^2 \bJ{\alpha}^{(1)}(r) + \dots\Big]\;.
\label{eq:Jexpansion}
\eeeq
In this letter we concentrate on the leading terms. The leading-order
contribution to the soft current is
\beq
\bJ{\mu}^{(0)}(r) = \sum_{k=1}^m \bT_k^r \frac{p_{k \mu}}{p_k\ldot p_r}
\,,
\label{eq:softJ0r}
\eeq
where $\bT_k^r \equiv \la c_r| T_k^{c_r}$ is the colour-charge operator
of parton $k$ if the emitted gluon has colour index $c_r$.
The amplitude is a colour-singlet state, therefore colour conservation
can be expressed as 
\beq
\sum_{k=1}^m \bT_k^r \ket{m}{}{(p_1,\dots,p_m)} = 0\,,
\label{eq:colorconserv}
\eeq
which implies conserved eikonal current,
\beq
p_r^\mu \bJ{\mu}^{(0)} (r) \ket{m}{}{(p_1,\dots,p_m)} = 
\sum_{k=1}^m \bT_k^r \ket{m}{}{(p_1,\dots,p_m)} = 0\,.
\label{eq:Jconserved}
\eeq

The soft limit of the squared matrix element can be obtained by squaring
\eqn{eq:SrM},
\beq
\bS{r} \SME{m+1}{0}{p_r,\dots} = 8\pi\as\mu^{2\ep}
\sum_{i=1}^m \sum_{k=1}^m \sum_{\rm pol.}
\eps_\mu(p_r) \eps_\nu^*(p_r) \frac12 \cS{ik}{\mu \nu}(r)
\SME{m; (i,k)}{0}{\dots}
\,,
\label{eq:SrM2}
\eeq
with
\beq
\cS{ik}{\mu \nu}(r) = \frac{4 p_i^\mu p_k^\nu}{s_{ir} s_{kr}}\,,
\label{eq:Sikmunu}
\eeq
where $s_{ir} = 2 p_i \ldot p_r$ and $\M{m;\,(i,k)}{(0)}{}$ 
denotes the colour-connected squared matrix element.
In evaluating the square, we use axial gauge with a light-like gauge
vector $n^\mu$ ($n^2 = 0$) to sum over the gluon polarizations, which
leads to the gluon polarization tensor $d^{\mu\nu}$:
\beq
d^{\mu\nu}(p_r, n) = \sum_{\rm pol.} \eps_\mu(p_r) \eps_\nu^*(p_r) =
-g^{\mu\nu} + \frac{p_r^\mu n^\nu + p_r^\nu n^\mu}{p_r\ldot n}
\,.
\label{eq:dmunu}
\eeq

As discussed above, in a physical gauge the collinear singularities are
due to the collinear splitting of an external parton, which remains true
in the soft limit.  The physical picture of collinear and soft
emissions suggests that the collinear part of the soft gluon emission
can be singled out by the diagonal term in the double sum ($i=k$) in
the soft limit of the squared amplitude. As mentioned,
colour-conservation implies conserved eikonal current (see
\eqn{eq:Jconserved}), therefore, the gauge terms in the gluon
polarization tensor do not contribute. However, if we want to separate
the diagonal terms from the rest, we must keep the gauge dependent
terms. A short calculation yields our new formula for the soft limit of
the squared matrix element:
\beeq
\bS{r} \SME{m+1}{0}{p_r,\dots} = -8\pi\as\mu^{2\ep}
\sum_{i=1}^m \aand\Bigg[
  \frac12 \sum_{k\ne i}^m \Bigg(\cS{ik}{}(r)
- \frac{2 s_{in}}{s_{rn} s_{ir}}
- \frac{2 s_{kn}}{s_{rn} s_{kr}}\Bigg)
\SME{m; (i,k)}{0}{\dots}
\nn \\&&
- \bT_i^2 \frac{2}{s_{ir}} \frac{s_{in}}{s_{rn}} \SME{m}{0}{\dots}
\Bigg]
\,,
\label{eq:SrM2new}
\eeeq
where $s_{in} = 2 p_i\ldot n$ and 
$\cS{ik}{}(r) = g_{\mu \nu} \cS{ik}{\mu \nu}(r)$, i.e. 
\beq
\calS_{ik}(r) = \frac{2 s_{ik}}{s_{ir}s_{kr}}\,.
\eeq
Exploiting
colour conservation, the gauge-dependent terms cancel and we obtain the
well-known form of soft-gluon factorization. It is easy to check that
the terms in the first line, containing the colour correlations, are
finite when parton $r$ is collinear to either $i$ or $k$, while those in
the second line become
\beq
\bC{ir}\bS{r} \SME{m+1}{0}{p_r,\dots} = 8\pi\as\mu^{2\ep}
\sum_{i=1}^m 
\bT_i^2 \frac{2}{s_{ir}} \frac{z_i}{z_r} \SME{m}{0}{\dots}
\,,
\label{eq:CirSr}
\eeq
in agreement with \Ref{Somogyi:2005xz}. Therefore, the soft-collinear 
contribution has been separated in the second line.

In order to define the soft limit in \eqn{eq:SrM2new} explicitly, we
have to fix the gauge vector $n^\mu$. The choice for the gauge vector
defines the infrared subtraction scheme employed for the computation of
radiative corrections because if we want to avoid the double counting of
the soft-collinear contribution, we have to be able to identify exactly
the same expression for the soft-collinear contribution both in the soft
limit (colour-diagonal piece in \eqn{eq:SrM2new}) and in the collinear
limit expressed in terms of momentum fractions (like in \eqn{eq:CirSr}).
The momentum fractions in the Sudakov parametrization of the collinear
limit are defined only in the strict collinear limit, and in order to
define subtraction terms one has to extend their definition over the
whole phase space. This can be done using a reference momentum $P^\mu$ as
\beq
z_i = \frac{s_{iP}}{s_{(ir)P}}\,,\qquad
z_r = \frac{s_{rP}}{s_{(ir)P}}\,,
\label{eq:ziwithP}
\eeq
where $s_{(ir)P} = s_{iP} + s_{rP}$  so that $z_i + z_r = 1$.
If one aims at setting up a subtraction scheme valid at any order in
perturbation theory, $P^\mu$ has to be chosen such that $s_{(ir)P}$
must not vanish in any of the multiple unresolved (soft and/or collinear)
regions of the phase space. In \Ref{Somogyi:2006cz} $P^\mu$  was chosen
to be the total incoming momentum $Q^\mu$ of the event, when $z_i/z_r =
s_{iQ}/s_{rQ}$, so \eqn{eq:CirSr} reads
\beq
\bC{ir}\bS{r} \SME{m+1}{0}{p_r,\dots} = 8\pi\as\mu^{2\ep}
\sum_{i=1}^m 
\bT_i^2 \frac{2}{s_{ir}} \frac{s_{iQ}}{s_{rQ}} \SME{m}{0}{\dots}
\,.
\label{eq:CirSrextended}
\eeq
This expression would be identical to the colour-diagonal piece in
\eqn{eq:SrM2new} if the gauge vector were chosen $n^\mu = Q^\mu$, which
is not possible in light-cone gauge. We may however, choose $n^\mu$
as
\beq
n^\mu = a_r(Q^\mu - b_r p_r^\mu)
\,,
\label{eq:nrmu}
\eeq
such that $n^2 = 0$ and $s_{rn} = s_{rQ} \equiv 2 p_r\ldot Q$. These
two requirements determine $a_r = 1$ and $b_r = Q^2/s_{rQ}$ uniquely.
This choice is equivalent to the Coulomb gauge in the center of mass
frame (rest frame of $Q^\mu$).  With this gauge vector and using
colour-conservation, we can rewrite \eqn{eq:SrM2new} as 
\beeq
\bS{r} \SME{m+1}{0}{p_r,\dots} = -8\pi\as\mu^{2\ep}
\sum_{i=1}^m \aand\Bigg[
  \frac12 \sum_{k\ne i}^m \Bigg(\cS{ik}{}(r)
- \frac{2 s_{iQ}}{s_{rQ} s_{ir}}
- \frac{2 s_{kQ}}{s_{rQ} s_{kr}}\Bigg)
\SME{m; (i,k)}{0}{\dots}
\nn \\&&
- \bT_i^2 \frac{2}{s_{ir}} \frac{s_{iQ}}{s_{rQ}} \SME{m}{0}{\dots}
\Bigg]
\,,
\label{eq:SrM2fin}
\eeeq
that is, formally we can make the simple replacement $n^\mu \to Q^\mu$.
Therefore, in order to
avoid double subtractions, we simply drop the
colour-diagonal terms from the soft factorization formula,
\beq
\bS{r} \SME{m+1}{0}{p_r,\dots} \to -8\pi\as\mu^{2\ep}
\sum_{i=1}^m 
  \frac12 \sum_{k\ne i}^m \Bigg(\cS{ik}{}(r)
- \frac{2 s_{iQ}}{s_{rQ} s_{ir}}
- \frac{2 s_{kQ}}{s_{rQ} s_{kr}}\Bigg)
\SME{m; (i,k)}{0}{\dots}\,.
\eeq
Using this purely soft limit, the collinear limit 
and the phase-space factorizations of \Ref{Somogyi:2006cz}, we arrive
at the same NLO subtraction scheme as in \Ref{Somogyi:2006cz}. Thus it
appears that our proposal for separating the soft and collinear limits
has not brought any advantage as compared to the usual technique of
subtracting both and adding back a proper soft-collinear compensation
term. Note however, that the separation of the soft and collinear
limits based on the colour structure makes the procedure very simple to
any order in perturbation theory.  Furthermore, it allows for defining
helicity-dependent subtractions and consequently, Monte Carlo summation
of helicities in the computation of radiative corrections to jet cross
sections. We discuss these two points in turn.

\section{Separation of soft and collinear singularities in multiple
infrared emissions}
\label{sec:NNLOseparation}

The overlapping structure of soft and collinear divergences in multiple
infrared emissions becomes very complex rapidly with increasing number
of unresolved partons. Already for two unresolved partons there are
triply overlapping regions. The separation of these requires very
careful analysis of the various infrared limits \cite{Somogyi:2005xz} 
and is a rather laborous excercise. The simple rules defined in the previous 
section can be applied to automate this cumbersome procedure as we discuss 
below.

We consider first the doubly soft-collinear limit when momenta
$p_i^\mu$ and $p_r^\mu$ are collinear and the gluon momentum
$p_s^\mu$, $(s\ne i,r)$ is soft.%
\footnote{We note that if parton $s$ is a fermion, then the squared
matrix element does not have a leading (doubly-unresolved) singularity
in this limit.} The factorization formula may be written in the form
\cite{Catani:1999ss}
\beeq
&&
\bSCS{ir;s}\SME{m+2}{0}{p_i,p_r,p_s,\ldots} =
(8\pi\as\mu^{2\ep})^2 \frac{1}{s_{ir}}
\label{eq:CSirs}\\
&&\times
\bra{m}{(0)}{(p_{(ir)},\ldots)}
\frac{1}{2}\left[\bJ{(ir)}^{\mu\dag}(s,\eps) d_{\mu\nu}(p_s,n)
\bJ{(ir)}^{\nu}(s,\eps)\right]\hP_{f_if_r}^{(0)}(z_i,z_r,k_{\perp};\eps)
\ket{m}{(0)}{(p_{(ir)},\ldots)}\,,
\nn
\eeeq
where the soft current $\bJ{(ir)}^\mu(s,\eps)$ is given by 
\eqns{eq:Jexpansion}{eq:softJ0r} with the replacement $r\to s$. The subscript
$(ir)$ serves simply to remind us that in the summation in \eqn{eq:softJ0r},
$k$ may take the value $(ir)$, in which case the summand is
\beq
\bT_{(ir)}^s\frac{p_{(ir) \mu}}{p_{(ir)}\ldot p_s} \equiv
(\bT_i^s+\bT_r^s)\frac{(p_i+p_r)_\mu}{(p_i+p_r)\ldot p_s}\,.
\eeq
Retracing the steps leading to \eqn{eq:SrM2fin}, in particular using colour 
conservation to make the replacement $n^\mu\to Q^\mu$, we find
\beeq
&&
\bSCS{ir;s}\SME{m+2}{0}{p_i,p_r,p_s,\ldots} =
-(8\pi\as\mu^{2\ep})^2 \frac{1}{s_{ir}}\Bigg[
\sum_{j=1}^{m}\sum_{k\ne j}^{m}\frac{1}{2}\left(\calS_{jk}(s)
-\frac{2s_{jQ}}{s_{sQ}s_{js}}-\frac{2s_{kQ}}{s_{sQ}s_{ks}}\right)
\nn\\&&\qquad\qquad\qquad\qquad\quad\times
\bra{m}{(0)}{(p_{(ir)},\ldots)}\hP_{f_if_r}^{(0)}(z_i,z_r,k_{\perp};\eps)
\bT_j \bT_k
\ket{m}{(0)}{(p_{(ir)},\ldots)}
\nn\\&&\qquad\qquad\quad
-\,\sum_{j\ne (ir)}^{m}\bT_j^2\frac{2}{s_{js}}\frac{s_{jQ}}{s_{sQ}}
\bra{m}{(0)}{(p_{(ir)},\ldots)}\hP_{f_if_r}^{(0)}(z_i,z_r,k_{\perp};\eps)
\ket{m}{(0)}{(p_{(ir)},\ldots)}
\nn\\&&\qquad\qquad\quad
-\,\bT_{(ir)}^2\frac{2}{s_{(ir)s}}\frac{s_{(ir)Q}}{s_{sQ}}
\bra{m}{(0)}{(p_{(ir)},\ldots)}\hP_{f_if_r}^{(0)}(z_i,z_r,k_{\perp};\eps)
\ket{m}{(0)}{(p_{(ir)},\ldots)}\Bigg]\,.
\eeeq
It is straightfroward that the product in the first two lines is finite
when $p_s^\mu$ is collinear to any other momentum that appears in the matrix
elements on the right hand side, including $p_{(ir)}^\mu$.\footnote{%
We remind the reader that unless explicitly indicated, we include parton $(ir)$
in the summations.} Furthermore, defining the momentum fractions as in 
\Ref{Somogyi:2006da}, we see that the third and fourth lines just reproduce the
double and triple collinear limits of the doubly soft-collinear factorization 
formula (Eqs.\ (4.36) and (4.32) in \Ref{Somogyi:2005xz}) respectively.

Next, we discuss the case of double soft parton emission.
The soft-gluon insertion rules are applicable in any order of
perturbation theory \cite{Catani:1999ss}, therefore, the soft-factorization 
formula for the amplitude can easily be given. For instance, for two soft 
partons the soft current has the expansion
\beq
\bJ{\alpha\beta}(r,s,\eps) = (\gs \mu^\ep)^2
\Big[\bJ{\alpha\beta}^{(0)}(r,s)
+ (\gs \mu^\ep)^2 \bJ{\alpha\beta}^{(1)}(r,s) + \dots\Big]\,,
\eeq
where the leading-order contribution in light-cone gauge is \cite{Catani:1999ss}
\beeq
\bJ{gg,\mu\nu}^{(0)}(r,s) \aand=
\sum_{i}^m \sum_{j\ne i}^m
  \bT_i^r\,\frac{p_{i \mu}}{p_i\ldot p_r}
\,\bT_j^s\,\frac{p_{j \nu}}{p_j\ldot p_s}
\nn\\ &&
+ \sum_{i}^m \left(
  \bT_i^r \,\bT_i^s\,\frac{p_{i \nu}}{p_i\ldot p_s}
\,\frac{p_{i \mu}}{p_i\ldot p_{(rs)}}
+ \bT_i^s \,\bT_i^r\,\frac{p_{i \mu}}{p_i\ldot p_r} 
\,\frac{p_{i \nu}}{p_i\ldot p_{(rs)}}\right)
\nn\\ &&
+ \sum_{i}^m
[\bT_i^r, \bT_i^s] \frac{p_{i \alpha}}{p_i\ldot p_{(rs)}} 
\,d^{\alpha\beta}(p_{(rs)}, n)
\frac{1}{p_r\ldot p_s} V_{\beta\mu\nu}(p_r, p_s)\,,
\label{eq:softJ0rsgg}
\\
\bJ{q\qb}^{(0)}(r,s) \aand=
\sum_{i}^m [\bT_i^r, \bT_i^s] \frac{p_{i \alpha}}{p_i\ldot p_{(rs)}} 
\,d^{\alpha\beta}(p_{(rs)}, n)
\frac{1}{p_r\ldot p_s}\gamma_\beta\,.
\label{eq:softJ0rsqq}
\eeeq
In \eqns{eq:softJ0rsgg}{eq:softJ0rsqq} $p_{(rs)}^\mu = p_r^\mu + p_s^\mu$ and 
\beq
V_{\beta\mu\nu}(p_r, p_s) =
\left[
  \frac12 (p_r-p_s)_\beta g_{\mu \nu}
+ p_{s\mu} g_{\nu \beta}
- p_{r\nu} g_{\mu \beta}
\right] \,.
\eeq
 
In order to separate the soft and collinear terms, we apply the same
procedure as in the case of single unresolved emission, which means
taking the square of the soft current and separating the colour-diagonal
terms. In the case of soft-$q\qb$ emission the algebra is relatively
simple. Using
$(\bT_i \ldot \bT_r)^\dag(\bT_k \ldot \bT_r) = \TR\,\bT_i \ldot \bT_k$,
we obtain
\beq
\bJ{q\qb}^{(0)\,\dag}(r,s) \pslash_r\pslash_s \bJ{q\qb}^{(0)}(r,s)=
\sum_i^m\sum_k^m \bT_i\ldot \bT_k
\,\frac{p_i^\mu\,d_{\mu\alpha}(p_{(rs)}, n)}{p_i\ldot p_{(rs)}}
\,\Pi_{q\qb}^{\alpha\beta}(p_r,p_s) 
\,\frac{d_{\beta\nu}(p_{(rs)}, n)\,p_k^\nu}{p_k\ldot p_{(rs)}}
\,,
\label{eq:JqqbJqqb}
\eeq
where $\Pi_{q\qb}^{\alpha\beta}(p_r,p_s)$ is the quark contribution to the
discontinuity of the gluon propagator,
\beq
\Pi_{q\qb}^{\alpha\beta}(p_r,p_s) = \frac{\TR}{(p_r\ldot p_s)^2}
\Big(p_r^\alpha p_s^\beta + p_s^\alpha p_r^\beta
    - g^{\alpha\beta} p_r\ldot p_s\Big)\,.
\eeq
Separating the colour-diagonal contributions we find
\beeq
&&
\bJ{q\qb}^{(0)\,\dag}(r,s) \pslash_r \pslash_s \bJ{q\qb}^{(0)}(r,s) =
\nn \\ &&\qquad
\sum_i^m\Bigg\{\sum_{k\ne i}^m \bT_i\ldot \bT_k
\,\frac{p_i^\mu\,d_{\mu\alpha}(p_{(rs)}, n)}{p_i\ldot p_{(rs)}}
\,\Pi_{q\qb}^{\alpha\beta}(p_r,p_s) 
\,\frac{d_{\beta\nu}(p_{(rs)}, n)\,p_k^\nu}{p_k\ldot p_{(rs)}}
\nn \\ &&\qquad\qquad
+ 4\,\bT_i^2\,\TR\,\frac{2}{s_{i(rs)} s_{rs}}
\left[\frac{s_{in}}{s_{(rs)n}}
- \frac{(s_{ir} s_{sn} - s_{is} s_{rn})^2}
{s_{i(rs)} s_{rs} s_{(rs)n}^2}\right]\Bigg\}
\,.
\label{eq:JqqbJqqbsep}
\eeeq
It is now straighforward to check that in the limit when the momenta
$p_r^\mu$, $p_s^\mu$ and either $p_i^\mu$ or $p_k^\mu$ are collinear, the
first line in \eqn{eq:JqqbJqqbsep} does not contain leading collinear
singularities, while the expression in the second line becomes
\beq
\bC{irs}\bJ{q\qb}^{(0)\,\dag}(r,s) \pslash_r \pslash_s \bJ{q\qb}^{(0)}(r,s) =
\sum_i^m
4\,\bT_i^2\,\TR\,\frac{2}{s_{i(rs)} s_{rs}}
\left[\frac{z_i}{z_r+z_s}
- \frac{(s_{ir} z_s - s_{is} z_r)^2}
{s_{i(rs)} s_{rs} (z_r+z_s)^2}\right]
\,,
\label{eq:CirsJqqbJqqbsep}
\eeq
which agrees with the expression in Eq.~(4.37) of \Ref{Somogyi:2005xz}. 

We choose the gauge vector similarly as in \eqn{eq:nrmu},
\beq
n^\mu = a_{rs} (Q^\mu - c_r p_r^\mu - c_s p_s^\mu)
\,.
\label{eq:nrsmu}
\eeq
Requiring $n^2 = 0$, $s_{rn}=s_{rQ}$ and $s_{sn}=s_{sQ}$ (so that also
$s_{(rs)n}=s_{(rs)Q}$), we can determine 
\beq
a_{rs} = \frac{1}{R}
\,,\qquad
c_r =  \frac{s_{sQ}}{s_{rs}}\,(1-R)
\,,\qquad
c_s =  \frac{s_{rQ}}{s_{rs}}\,(1-R)
\,,\qquad
R = \sqrt{1 - \frac{Q^2 s_{rs}}{s_{rQ} s_{sQ}}}
\,.
\eeq
We can now use colour-conservation to see that the
formal substitution $n^\mu \to Q^\mu$ can again be applied to obtain
\beeq
\bJ{q\qb}^{(0)\,\dag}(r,s) \pslash_r \pslash_s \bJ{q\qb}^{(0)}(r,s) =
\sum_i^m\aand\Bigg\{\sum_{k\ne i}^m \bT_i\ldot \bT_k
\,\frac{p_i^\mu\,d_{\mu\alpha}(p_{(rs)}, Q)}{p_i\ldot p_{(rs)}}
\,\Pi_{q\qb}^{\alpha\beta}(p_r,p_s) 
\,\frac{d_{\beta\nu}(p_{(rs)}, Q)\,p_k^\nu}{p_k\ldot p_{(rs)}}
\nn \\ &&
+ 4\,\bT_i^2\,\TR\,\frac{2}{s_{i(rs)} s_{rs}}
\left[\frac{s_{iQ}}{s_{(rs)Q}}
- \frac{(s_{ir} s_{sQ} - s_{is} s_{rQ})^2}
{s_{i(rs)} s_{rs} s_{(rs)Q}^2}\right]\Bigg\}
\,.
\label{eq:JqqbJqqbfin}
\eeeq
The soft-collinear term is now separated in the term proportional to
$\bT_i^2$. This term can simply be dropped because it also appears in the
Altarelli--Parisi splitting function of the $q \to q_i \qb_r' q_s'$
collinear splitting \cite{Catani:1999ss} if the momentum fractions are 
defined as in \Ref{Somogyi:2006da}.

The same procedure works also when two soft gluons are emitted.
In this case it is more convenient to rewrite the double-current
in an equivalent form in terms of colour-charge anticommutators and
commutators:
\beq
\bJ{gg,\mu\nu}^{(0)}(r,s) =
\sum_i^m\sum_j^m \frac12 \{\bT_i^r, \bT_j^s\}\,A_{\mu\nu}^{ij}(p_r,p_s)
+ \sum_{i}^m [\bT_i^r,\bT_i^s]\,C_{\mu\nu}^{i}(p_r,p_s)
\,,
\label{eq:softJ0rs2}
\eeq
where 
\beq
A_{\mu\nu}^{ij}(p_r,p_s) =
\frac{p_{i\mu}}{p_i\ldot p_r}\,\frac{p_{j\nu}}{p_j\ldot p_s}
\eeq
and
\beq
C_{\mu\nu}^{i}(p_r,p_s) =
\frac{p_{i \alpha}d^{\alpha\beta}(p_{(rs)},n)\,V_{\beta\mu\nu}(p_r,p_s)}
     {p_r\ldot p_s\,p_i\ldot p_{(rs)}}
- \frac12 \frac{p_i\ldot (p_r-p_s)}{p_i\ldot (p_r+p_s)} 
  \frac{p_{i\mu} p_{i\nu}}{p_i\ldot p_r\,p_i\ldot p_s}
\,.
\eeq
In evaluating the square, we use the following colour identities:
\beeq
\frac14 \{\bT_i^r,\bT_j^s\}^\dag \{\bT_k^r,\bT_l^s\} \aand=
  \frac12 \{\bT_j\ldot \bT_l,\bT_i\ldot \bT_k\}
+ \frac14 \CA\,\bT_i\ldot \bT_k\,\delta_{ij}\,\delta_{kl}
\nn\\ &&
+ \frac12 \CA\Big[
\bT_i\ldot \bT_k\,\delta_{il}
- \bT_i\ldot \bT_l\,(\delta_{ik}+\delta_{kl})
\Big]\,\delta_{jk}
\nn\\ &&
+ \frac{\ri}{2} \:f_{abc}
( \bT_j^a\,\bT_i^b\,\bT_k^c \,\delta_{il}
- \bT_l^a\,\bT_k^b\,\bT_i^c \,\delta_{jk} )
\,,
\label{eq:++}
\\[3mm]
  \frac12 \{\bT_i^r,\bT_j^s\}^\dag [\bT_k^r,\bT_k^s]
\aand+\:\frac12 [\bT_k^r,\bT_k^s]^\dag \{\bT_i^r,\bT_j^s\} =
\CA\,\bT_i\ldot\bT_j (\delta_{ik} - \delta_{jk})
\,,
\label{eq:+-plus-+}
\\[3mm]
[\bT_i^r,\bT_i^s]^\dag \,[\bT_j^r,\bT_j^s] \aand= \CA\,\bT_i \ldot \bT_j
\,.
\label{eq:--}
\eeeq
The terms proportional to the structure constants $f_{abc}$ in
\eqn{eq:++} are antisymmetric, while the kinematic factors that these
terms multiply are symmetric when $i$ with $k$ and $j$ with $l$ are
simultaneously interchanged. Therefore, these do not contribute to
the square. The remaining terms contain four-fold and two-fold
summations. That with four summations has single diagonal
($\sum_i \bT_i^2 \sum_j \sum_{l\ne j} \bT_j \ldot \bT_l$) and double diagonal
($\sum_i \bT_i^2 \sum_{j\ne i} \bT_j^2$ or $\sum_i \bT_i^2 \bT_i^2$) terms,
which separate the soft-collinear (Eq.~(4.45) in \Ref{Somogyi:2005xz}),
doubly-collinear (Eq.~(4.43) in \Ref{Somogyi:2005xz}) and abelian
triply-collinear (Eq.~(4.40) in \Ref{Somogyi:2005xz}) contributions in
the doubly-soft emissions, respectively. The contribution with two
summations has single diagonal terms ($\sum_i \bT_i^2$), which separate
the non-abelian triply-collinear pieces (Eq.~(4.41) in
\Ref{Somogyi:2005xz}). Again, the colour non-diagonal pieces do not
have leading singularities in the various collinear limits.
We choose the gauge vector as in \eqn{eq:nrsmu}.

\section{Monte Carlo treatment of helicity summation in NLO computations}
\label{sec:MC_helicity}

In \Refs{Draggiotis:1998gr,Draggiotis:2002hm} a Monte Carlo integration over 
a phase variable for reproducing
the helicity sums in the squared matrix element was introduced in order
to save CPU time when computing multijet cross sections. While this
approach has been found useful for computing cross sections at the
LO accuracy, its extension to NLO computations is hampered by the
explicit summation over the helicities in the subtraction terms used in
any of the known NLO calculations. With the separation of singularities
presented here, the Monte Carlo treatment of the helicity summation
becomes possible by keeping the helicity states for the unresolved
partons. 

The Monte Carlo treatment of the helicity summation requires
helicity-dependent subtraction terms. Although the soft and collinear
limits of helicity amplitudes are well-known \cite{Mangano:1990by}, the overlap between
these cannot be determined at the amplitude level. In taking the square
of the amplitude, keeping the helicity-dependence in the collinear
subtractions is straightforward. In order to keep the helicity-dependence
in the soft subtractions, one needs soft terms from which the collinear
singularities are subtracted in a helicity-independent way. Our recipe
does precisely that: the soft-collinear terms are identified in the
colour sum. In order to define helicity-dependent soft subtractions, we
do not substitute the polarization tensor $d_{\mu\nu}$ for the
summations over gluon polarizations. The collinear contributions can
nevertheless be separated in the colour-diagonal terms as discussed in
\sect{sec:softsep_NLO}.

We start from the soft factorization formula \eqn{eq:SrM2}, but now we
do not sum over the polarizatons of gluon $r$ or the helicities of the
other $m$ partons.  Separating the colour-diagonal terms as before, we have
\beeq
\bS{r} \SME{m+1}{0}{p_r^{\lambda},\dots} = 8\pi\as\mu^{2\ep}
\sum_{i=1}^m \aand\Bigg[\sum_{k\ne i}^m
\eps_\mu^\lambda(p_r, n)
\,\frac12 \cS{ik}{\mu \nu}(r)
\,\eps_\nu^{-\lambda}(p_r, n)
\SME{m; (i,k)}{0}{\dots}
\nn \\ &&
+ \bT_i^2\,\eps_\mu^\lambda(p_r, n)
\,\frac12 \cS{ii}{\mu \nu}(r)
\,\eps_\nu^{-\lambda}(p_r, n)
\SME{m}{0}{\dots}\Bigg]
\,.
\qquad~
\label{eq:SrM2pol}
\eeeq
Next, notice that the double summation over $i$ and $k$ in the first line
of \eqn{eq:SrM2pol}
effectively symmetrizes $\calS^{\mu\nu}_{ik}(r)$ in its Lorentz indices, while
$\calS^{\mu\nu}_{ii}(r)$ in the second line is already symmetric in $\mu$ and $\nu$.
Thus only the symmetric part of $\eps_\mu^\lambda(p_r, n)\eps_\nu^{-\lambda}(p_r, n)$
contributes. However
\beq
\eps_{(\mu}^\lambda(p_r, n)\eps_{\nu)}^{-\lambda}(p_r, n) \equiv
\frac{1}{2}\left(\eps_\mu^\lambda(p_r, n)\eps_\nu^{-\lambda}(p_r, n) +
\eps_\nu^\lambda(p_r, n)\eps_\mu^{-\lambda}(p_r, n)\right) = 
\frac{1}{2}d_{\mu\nu}(p_r,n)\,,
\eeq
and so by exactly the same steps that lead to \eqn{eq:SrM2fin}, we obtain
\beeq
\bS{r} \SME{m+1}{0}{p_r^{\lambda},\dots} =
-4\pi\as\mu^{2\ep}\sum_{i=1}^m \aand\Bigg[
  \frac12 \sum_{k\ne i}^m \Bigg(\cS{ik}{}(r)
- \frac{2 s_{iQ}}{s_{rQ} s_{ir}}
- \frac{2 s_{kQ}}{s_{rQ} s_{kr}}\Bigg)
\SME{m; (i,k)}{0}{\dots}
\nn \\&& 
- \bT_i^2 \frac{2}{s_{ir}} \frac{s_{iQ}}{s_{rQ}} \SME{m}{0}{\dots}
\Bigg]
\,.
\label{eq:SrM2polfin}
\eeeq
\eqn{eq:SrM2polfin} shows that the soft factorization formula is
independent of the polarization of the emitted soft gluon.
Performing the summation over the helecity
$\lambda$ of the soft gluon (a multiplication by two) as well as the
helicities of the rest of the partons, we trivially recover the
expression in \eqn{eq:SrM2fin}. Dropping the soft-collinear term in the
second line, we find the helicity-dependent%
\footnote{Recall that in \eqn{eq:SrM2polfin}, the helicities of all
partons (not just the helicity $\lambda$ of the soft gluon) are fixed
and not summed over.}
purely soft subtraction term, that clearly does not contain leading
singularities when parton $r$ is collinear to either parton $i$ or
parton $k$.  The helicity-dependent collinear subtraction terms,
including the soft-collinear contributions, can be obtained by squaring
the collinear factorization formulae for helicity-amplitudes
\cite{Mangano:1990by,Birthwright:2005vi}.  Helicity-dependent
subtraction terms may also be defined using the antennae factorization
expressions of \Refs{Kosower:1997zr,Kosower:2002su}.

We note that we have checked the validity of \eqn{eq:SrM2polfin} explicitly
for the processes $e^+e^- \to q\qb g$ and $e^+e^- \to q\qb gg$ starting from
the expressions for the relevant helicity amplitudes as given in \Ref{Bern:1997sc}.

\section{Summary}

We have defined a new method for separating the soft and the collinear
singularities in the QCD factorization formulae. The rules of the method
are very simple. One starts with the soft-gluon insertion rules for
finding the soft limit of the amplitudes. In taking the square of those
we do not exploit colour-conservation for cancelling the gauge terms
that appear in the physical polarisations of the soft gluon, rather we
separate the colour-diagonal contributions. The colour non-diagonal
contributions are free of leading collinear singularities, while the
collinear limit of the diagonal contributions lead to the known
singular expressions.  

For the gauge vector $n^\mu$ we may choose a light-like vector whose
space-like component points into opposite to that of the unresolved
gluon $r$ and require that $s_{rn} = s_{rQ}$, which ensures that we can
perform the formal substitution $n^\mu \to Q^\mu$, where $Q^\mu$ is the
total four-momentum of the event. This amounts to using Coulomb gauge.
We have shown that this gauge can be easily generalized to any order in
perturbation theory.  Choosing the momentum fractions in the collinear
subtractions as in \Ref{Somogyi:2006cz}, the terms separated in the
colour-diagonal contributions can be identified also in the collinear
subtractions, therefore, can be dropped. Other choices for the gauge
vectors are also possible, but we do not discuss that further in this
letter.

This technique can be automatized easily and applied in any order of
perturbation theory. It also facilitates the use of Monte Carlo helicity
summation in the computation of the radiative corrections, therefore, can
lead to significant reduction of CPU time when there are many partons in
the final state, which is the most interesting case for the LHC. We
should mention that a Monte Carlo treatment of colour summation also
results in significant reduction of CPU time. Such treatment of colour in
NLO computations is also facilitated by this new method.

Our method could be useful also for improving parton shower Monte Carlo
algorithms. The algorithms that are implemented presently treat the
colour correlations approximately in the large $N_{c}$ limit and 
sum up the leading and the next-to-leading logarithms at leading-colour
accuracy. The summation of the subleading logarithms with exact colour
treatment is rather difficult. However, one can go beyond the
leading-colour approximation systematically by considering the
subleading colour contributions pertubatively instead of exponentiating
them. To do this we have to introduce subtraction terms for the
subleading colour contributions in the parton shower algorithm.  The
technique presented here can be used to define splitting kernels and
the corresponding counterterms for a parton shower algorithm since the
separation of the singularities is governed by the colour structure,
which provides a good control over the large logarithms and the colour
structure simultaneously. We shall elaborate these ideas in separate
publications.  

\section*{Acknowledgments}
This research was supported by the Hungarian Scientific Research Fund
grant OTKA K-60432. Z.N. is grateful to D. Soper, while Z.T. to T.
Gehrmann for useful discussions.

\end{document}